----------
X-Sun-Data-Type: default
X-Sun-Data-Description: default
X-Sun-Data-Name: qcd11.tex
X-Sun-Content-Lines: 251

\documentstyle[preprint,aps]{revtex}
\begin{document}
\draft

\title {QCD Scales in Finite Nuclei}
\author {J. L. Friar and D. G. Madland}
\address {Theoretical Division, Los Alamos National Laboratory, Los Alamos,
          New Mexico, 87545}
\author {B. W. Lynn}
\address {Clarendon Laboratories, Oxford University, Oxford OX1 3PU, Great
Britain}
\maketitle

\begin{abstract}
The role of QCD scales and chiral symmetry in finite nuclei is examined. The
Dirac-Hartree mean-field coupling constants of Nikolaus, Hoch, and Madland
(NHM) are scaled in accordance with the QCD-based prescription of Manohar and
Georgi. Whereas the nine empirically-based coupling constants of NHM span
thirteen
orders of magnitude, the scaled coupling constants are almost all {\it
natural},
being dimensionless numbers of order one. We speculate that this result
provides
good evidence that QCD and chiral symmetry apply to finite nuclei.
\end{abstract}

\narrowtext

Although QCD is widely believed to be the underlying theory of the strong
interaction, a direct description of nuclear properties in terms of the {\it
natural} degrees of freedom of that theory, quarks and gluons, has proven
elusive. The problem is that at sufficiently low energy, the {\it physical}
degrees of freedom of nuclei are nucleons and (intranuclear) pions.
Nevertheless, QCD can be mapped onto the latter Hilbert space and the resulting
effective field theory is capable in principle of providing a dynamical
framework for nuclear calculations. This framework is usually called chiral
perturbation theory ($\chi$PT).

Two organizing principles govern this $\chi$PT: (1) (broken) chiral symmetry
(which is manifest in QCD) and (2) an expansion in powers of $(Q/\Lambda)$,
where
$Q$ is a general intranuclear momentum or pion mass, and $\Lambda$ is the
generic
QCD large-mass scale $\sim$1 GeV, which in a loose sense indicates the
transition region between the two alternative sets of degrees of freedom
indicated above (that is, quark-gluon versus nucleon-pion). Typically, one
constructs Lagrangians (that is, interactions) that display (broken) chiral
symmetry and retains only those terms with exponents less than or equal to some
fixed power of (1/$\Lambda$). The chiral symmetry itself provides a crucial
constraint: a general term has the structure $\sim$ $(Q/\Lambda)^{N}$ and $N
\geq 0$ is mandated. This guarantees that higher-order constructions in
perturbation theory ({\it viz.}, loops) will have even higher (not lower)
powers of $(Q/\Lambda)$. The price one pays for this mapping from {\it natural}
to {\it
effective} degrees of freedom is an infinite series of interaction terms, where
coefficients are unknown and must be determined from experiment.

To date only a few nuclear calculations have been performed within this
framework. The seminal work of Weinberg \cite{W90} highlighted the role of
power counting and chiral symmetry in weakening N-body forces. That is,
two-nucleon forces are stronger than three-nucleon forces, which are stronger
than four-nucleon forces, $\cdots$ . This chain makes nuclear physics
tractable. Van Kolck and collaborators \cite{vk1} developed a nuclear potential
model, including one-loop (two-pion exchange) contributions. Friar and Coon
\cite{fc} developed non-adiabatic two-pion-exchange forces, while van Kolck,
Friar and Goldman \cite{vfg} examined isospin violation in the nuclear force.
Rho, Park, and Min \cite{rho} were the first to treat external electromagnetic
and weak interactions with nuclei. Essentially all of this work was focused on
few-nucleon systems, where computational techniques are sophisticated. Only the
work of Lynn \cite{L93} on (nuclear) chiral liquids was specifically directed
at heavier nuclei and, more recently, Gelmini and Ritzi \cite{GR95} have
calculated nuclear matter properties using lowest order nonlinear chiral
effective Lagrangians.

Is there any evidence for chiral symmetry or QCD scales in finite nuclei? The
tractability and astonishing success of the recent few-nucleon calculations of
$^2$H, $^3$H, $^3$He, $^4$He, $^5$He, $^6$He, $^6$Li, and $^6$Be with only a
weak three-nucleon force and no four-nucleon force confirms Weinberg's
power-counting prediction \cite{W90} and yields strong but indirect evidence
for chiral symmetry. The work of Lynn\cite{L93} established a procedure for
going beyond few-nucleon systems. Nuclear (N-body) forces either have zero
range
or are generated by pion exchange. Following Manohar and Georgi \cite{MG84} we
can scale a generic Lagrangian component as
\begin{equation}
{\cal L} \sim -c_{l m n}
\biggl[ \frac{\overline{\psi}\psi}{f^2_{\pi} \Lambda} \biggr]^l
\biggl[ \frac{\vec{\pi}}{f_{\pi}} \biggr]^{m}
\biggl[ \frac{\partial^{\mu}, m_{\pi}}{\Lambda} \biggr]^n
f^2_{\pi} \, \Lambda^2 \,
\label{eq:1}
\end{equation}
where $\psi$ and $\vec\pi$ are nucleon and pion fields, respectively, $f_{\pi}$
and $m_{\pi}$ are the pion decay constant, 92.5 MeV, and pion mass, 139.6 MeV,
respectively, $\Lambda \sim 1$ GeV has been discussed above, and
($\partial^\mu$, $m_\pi$) signifies either a derivative or a power of the pion
mass. Dirac matrices and isospin operators (we use $\vec{t}$ here rather than
$\vec{\tau}$) have been ignored. Chiral symmetry demands \cite{W79}
\begin{equation}
\Delta = l + n - 2 \geq 0 \: .
\label{eq:2}
\end{equation}
Thus the series contains only {\it positive} powers of (1/$\Lambda$). If the
theory is
{\it natural}\cite{L93,MG84}, the dimensionless coefficients $c_{lmn}$ are of
order (1). Thus, all information on scales ultimately resides in the $c_{lmn}$.
If they are natural, scaling works. Our limited experience with nuclear-force
models suggests that natural coefficients are the rule.

Unfortunately, zero-range nuclear-force models are not widely used. However, a
recent calculation has been performed using zero-range forces for an extended
range of mass number A and this work provides significant new information on
QCD and chiral symmetry in nuclei. Nikolaus, Hoch, and Madland (NHM)
\cite{NHM92} used a series of zero-range interactions to perform Dirac-Hartree
calculations in mean-field approximation for a total of fifty-seven nuclei.
Their Lagrangian [using their notation] is given by
\begin{equation}
{\cal L} = {\cal L}_{free} + {\cal L}_{4f} + {\cal L}_{hot} + {\cal L}_{der}
+ {\cal L}_{em} \ ,
\label{eq:3}
\end{equation}
where ${\cal L}_{free}$ and ${\cal L}_{em}$ are the kinetic and electromagnetic
terms, respectively, and

\begin{eqnarray}
{\cal L}_{4f} & = & -\frac{1}{2}{{\alpha}_{S}}({\bar{\psi}}{\psi})({\bar{\psi}}
{\psi})-\frac{1}{2}{{\alpha}_{V}}({\bar{\psi}}{{\gamma}_{\mu}}{\psi})
({\bar{\psi}}{{\gamma}^{\mu}}{\psi}) \nonumber \\
&   & -\frac{1}{2}{{\alpha}_{TS}}({\bar{\psi}}{\vec{\tau}}{\psi}){\cdot}
({\bar{\psi}}{\vec{\tau}}{\psi})
-\frac{1}{2}{{\alpha}_{TV}}({\bar{\psi}}{\vec{\tau}}{{\gamma}_{\mu}}{\psi})
{\cdot}({\bar{\psi}}{\vec{\tau}}{{\gamma}^{\mu}}{\psi}) \ , \label{eq:4} \\
 & &  \nonumber \\
{\cal L}_{hot} & = & -\frac{1}{3}{{\beta}_{S}}({\bar{\psi}}{\psi})^{3}
-\frac{1}{4}{{\gamma}_{S}}({\bar{\psi}}{\psi})^{4}
 -\frac{1}{4}{{\gamma}_{V}}[({\bar{\psi}}{{\gamma}_{\mu}}{\psi})
({\bar{\psi}}{{\gamma}^{\mu}}{\psi})]^{2} \ , \ {\rm and} \label{eq:5} \\
 & &  \nonumber \\
{\cal L}_{der} & = & -\frac{1}{2}{{\delta}_{S}}({\partial_{\nu}}
{\bar{\psi}}\psi)({\partial^{\nu}}{\bar{\psi}}\psi)
  -\frac{1}{2}{{\delta}_{V}}({\partial_{\nu}}{\bar{\psi}}{\gamma_{\mu}}\psi)
({\partial^{\nu}}{\bar{\psi}}{\gamma^{\mu}}\psi) \label{eq:6} \ .
\end{eqnarray}

\noindent In these equations, $\psi$ is the nucleon field, the subscripts $S$
and $V$ refer to the isoscalar-scalar and isoscalar-vector densities,
respectively, and the subscripts $TS$ and $TV$ refer to the isovector-scalar
and isovector-vector densities, respectively, containing the nucleon isospin
operator $\vec{\tau}$. The nine coupling constants of the NHM Lagrangian were
determined in a self-consistent procedure that solved the model equations for
several nuclei simultaneously in a nonlinear least-squares adjustment algorithm
with respect to measured ground-state observables (Table IV of Ref.
\cite{NHM92}). The predictive power of the
extracted coupling constants is quite good both for other finite nuclei and for
the properties of saturated nuclear matter (see Tables VIII, IX, and XI
of Ref. \cite{NHM92}).

${\cal L}_{4f}$ contains four two--nucleon--force terms corresponding to
$\Delta = 0$, the first term of ${\cal L}_{hot}$ is a three--nucleon--force
term corresponding to $\Delta = 1$, whereas the remaining two terms are
four--nucleon--force terms corresponding to $\Delta = 2$. Finally, ${\cal
L}_{der}$ contains two nonlocal two--nucleon--force terms, also corresponding
to $\Delta = 2$. The derivative terms act on ${\bar{\psi}}\psi$, rather than on
just one of the fields, because the latter generate a factor $E \cong M$, the
nucleon mass, whereas the former generate an energy difference that is
considerably smaller. The latter terms would spoil the series in Eq. (1) since
$M \cong \Lambda$. However, either by a transformation or by rearranging the
series, this problem could in principle be eliminated \cite{L93}.

The construction of the NHM Lagrangian was motivated by empirically-based
improvements to a Walecka type scalar-vector model \cite{SW79,HS81}, but using
contact (zero-range) interactions to allow treatment of the Fock (exchange)
terms. It was not motivated either by power counting or by chiral symmetry. The
pion degrees of freedom are ignored and the Lagrangian is not complete;
additional operators in each order of (1/$\Lambda)^{\Delta}$ are possible.
Specifically, the NHM Lagrangian, Eqs. (4)--(6), has four operators in order
(1/$\Lambda)^{0}$, one operator in order (1/$\Lambda)^{1}$, and four operators
in order (1/$\Lambda)^{2}$, constituting an incomplete mix of three different
orders in (1/$\Lambda)$.

Nevertheless, a meaningful comparison can be made of the generic chiral
Lagrangian given by Eqs.(1) and (2) and the NHM Lagrangian given by
Eqs.(4)--(6), precisely because our test of naturalness does not care whether a
specific $c_{lmn}$ coefficient is 0.5 or 2.0. Changing (refining) the model by
adding terms would change {\it all} of the $c_{lmn}$, but the same test of
naturalness still applies. Adding new terms would simply change a specific
coefficient by an amount $\sim$ 1 (or less).

The nine coupling constants of the NHM Lagrangian are shown in Table 1, both in
dimensional and dimensionless form [the latter obtained by equating Eqs.(1) and
(4)--(6), with $\Lambda = 1$ GeV, using isospin operators $\vec{t}$ in Eq.(1),
and solving for $c_{lmn}$ in terms of $\alpha$, $\beta$, $\gamma$, and
$\delta$]. In the former form they span more than thirteen orders of magnitude,
while in the latter form six of the nine coupling constants can be regarded as
natural. Only the very small $\alpha_{TS}$ and large $\gamma_{S}$ and
$\gamma_{V}$ are unnatural. However, the sum of the latter appears to be
natural, and we speculate that the difference may not be well determined in the
least-squares adjustments to the measured observables. The unnaturally small
$\alpha_{TS}$, if correct, would presuppose a symmetry to preserve its small
value.

Although these results were not obtained as a test of chiral symmetry and QCD
scales (NHM at that time were unaware of these developments) and hence are
imperfect, they are conversely completely unbiased. This result is very
indicative of the role of chiral symmetry and QCD in finite nuclei. A
systematic study of this approach is clearly indicated.

\newpage
\begin{table}[t]
\Large
\caption{ Optimized Coupling Constants for the NHM Lagrangian
and Corresponding Dimensional Power Counting Coefficients
and Chiral Expansion Order}
\begin{tabular}{crcrl}
\vspace{18pt}
Coup. Const. & Magnitude & Dimension & $c_{lmn}$ & Order \\ \hline
$\alpha_{S}$ & -4.508${\times}10^{-4}$ & MeV$^{-2}$ & -1.93 & $\Lambda^{0}$ \\
$\alpha_{TS}$ & 7.403${\times}10^{-7}$ & MeV$^{-2}$ & 0.013 & $\Lambda^{0}$ \\
$\alpha_{V}$ & 3.427${\times}10^{-4}$ & MeV$^{-2}$ & 1.47 & $\Lambda^{0}$ \\
$\alpha_{TV}$ & 3.257${\times}10^{-5}$ & MeV$^{-2}$ & 0.56 & $\Lambda^{0}$ \\
$\beta_{S}$ & 1.110${\times}10^{-11}$ & MeV$^{-5}$ & 0.27 & $\Lambda^{-1}$ \\
$\gamma_{S}$ & 5.735${\times}10^{-17}$ & MeV$^{-8}$ & 8.98 & $\Lambda^{-2}$ \\
$\gamma_{V}$ & -4.389${\times}10^{-17}$ & MeV$^{-8}$ & -6.87 & $\Lambda^{-2}$
\\
$\delta_{S}$ & -4.239${\times}10^{-10}$ & MeV$^{-4}$ & -1.81 & $\Lambda^{-2}$
\\
$\delta_{V}$ & -1.144${\times}10^{-10}$ & MeV$^{-4}$ & -0.49 & $\Lambda^{-2}$
\end{tabular}
\end{table}

\begin{thebibliography}{999}
\bibitem{W90}
S.\ Weinberg, Phys.\ Lett.\ {\bf 251B}, 288 (1990).
\bibitem{vk1}  C.\ Ord\'o\~nez and U.\ van Kolck, Phys.\ Lett.\ {\bf B291}, 459
(1992); C.\ Ord\'o\~nez, L.\ Ray, and U.\ van Kolck, Phys.\ Rev.\
Lett.\ {\bf 72}, 1982 (1994).
\bibitem{fc} J.\ L.\ Friar and S.\ A.\ Coon, Phys.\ Rev.\ C {\bf 49}, 1272
(1994); S.\ A.\ Coon and J.\ L.\ Friar, Phys.\ Rev.\ C {\bf 34}, 1060 (1986).
\bibitem{vfg} U.\ van Kolck, J.\ L.\ Friar, and T.\ Goldman,
Phys.\ Lett.\ (submitted).
\bibitem{rho}  T.-S.\ Park, D.-P.\ Min, and M.\ Rho, Phys.\ Rev.\ Lett.\
{\bf 74}, 4153 (1995); M.\ Rho, Phys.\ Rev.\ Lett.\ {\bf 66}, 1275 (1991);
T.-S.\ Park, D.-P.\ Min, and M.\ Rho, Phys.\ Rep. {\bf 233}, 341 (1993).
\bibitem{L93}
B.\ W.\ Lynn, Nucl.\ Phys.\ {\bf B402}, 281 (1993).
\bibitem{GR95}
G.\ R.\ Gelmini and B.\ Ritzi, Phys.\ Lett.\ B {\bf 357}, 431 (1995).
\bibitem{MG84}
A.\ Manohar and H.\ Georgi, Nucl.\ Phys.\ {\bf B234}, 189 (1984).
\bibitem{W79}
S.\ Weinberg, Physica {\bf 96A}, 327 (1979).
\bibitem{NHM92}
B.\ A.\ Nikolaus, T.\ Hoch, and D.\ G.\ Madland, Phys.\ Rev.\ C {\bf 46},
1757 (1992).
\bibitem{SW79}
B.\ D.\ Serot and J.\ D.\ Walecka, Phys.\ Lett.\ {\bf 87B}, 172 (1979).
\bibitem{HS81}
C.\ J.\ Horowitz and B.\ D.\ Serot, Nucl.\ Phys.\ {\bf A368}, 503 (1981).

\end{thebibliography}
\end{document}